\documentclass[useAMS,usenatbib]{mn2e}
\usepackage{psfig}

\title[Eccentricity of HLX-1]{Eccentricity of HLX-1}
\author[R. Soria]{Roberto Soria$^{1}$\thanks{E-mail:
roberto.soria@icrar.org}\\  
$^{1}$International Centre for Radio Astronomy Research, Curtin University, 
GPO Box U1987, Perth, WA 6845, Australia}
\begin{document}

\date{Accepted 2012 October 15; Received 2012 October 4; 
in original form 2012 August 15}

\pagerange{\pageref{firstpage}--\pageref{lastpage}} \pubyear{2011}

\maketitle

\label{firstpage}

\begin{abstract}
We compare the outer radius of the accretion disc in the intermediate-mass 
black hole candidate HLX-1 as estimated from the UV/optical continuum, 
with the values estimated from its outburst decline timescales. We fit  
the {\it Swift} 2010 outburst decline lightcurve with an exponential 
decay, a knee and a linear decay. We find that the disk has an outer radius 
$10^{12}$ cm $\la R_{\rm out} \la 10^{13}$ cm, only an order of magnitude 
larger than typical accretion discs in the high/soft state of Galactic 
black holes. By contrast, the semimajor axis is $\approx$ a few $10^{14}$ cm.
This discrepancy can be explained with a highly eccentric orbit.
We estimate the tidal truncation radius and circularization radius 
around the black hole at periastron, and impose that they are similar 
or smaller than the outer disk radius. We obtain that $e \ga 0.95$, 
that the radius of the donor star is $\la$ a few solar radii, 
and that the donor star is not at risk of tidal disruption.
If the companion star fills 
its Roche lobe and impulsively transfers mass only around periastron, 
secular evolution of the orbit is expected to increase 
eccentricity and semimajor axis even further.
We speculate that such extremely eccentric systems may have the same 
origin as the S stars in the Galactic centre.


\end{abstract}


\section{Introduction}

The point-like X-ray source 2XMM\,J011028.1$-$460421 (henceforth, 
HLX-1 for simplicity) is the strongest intermediate-mass black hole 
(IMBH) candidate known to date \citep{far09,wie10,dav11,ser11}.
It is seen in the sky at a distance of $\approx 8\arcsec$ 
from the nucleus of the S0 galaxy ESO\,243-49 (redshift $z = 0.0224$, 
luminosity distance $\approx 95$ Mpc, distance modulus $\approx 34.89$ mag; 
at this distance, $1\arcsec \approx 460$ pc).
Its X-ray luminosity and spectral variability \citep{far09,god09,ser11} 
and its radio flares detected in association with the X-ray 
outbursts \citep{web12} are consistent with the canonical 
state transitions and jet properties of an accreting BH. 
With a peak X-ray luminosity $\approx 10^{42}$ erg s$^{-1}$, 
the BH mass required to be consistent 
with the Eddington limit is $\sim 10^4 M_{\odot}$. A similar value 
is obtained from spectral modelling of the thermal X-ray component, 
which is consistent with emission from an accretion disc 
\citep{far09,dav11,ser11}.
If these BH mass estimates are correct, HLX-1 is way too massive 
to have been formed from any stellar evolution process. A more likely 
scenario is that it is the nuclear BH (perhaps still surrounded 
by its own nuclear star cluster) of a disrupted 
dwarf satellite galaxy, accreted by ESO\,243-49 \citep{map12,kin05}.
HLX-1 has a point-like, blue optical counterpart 
($B \sim V \sim 24$ mag near the outburst peak; \citealt{far12,sor12,sor10}).
The presence of H$\alpha$ emission at a redshift consistent with that 
of ESO\,243-49 \citep{wie10} is perhaps the strongest argument for 
a true physical association. It is still debated whether the optical 
continuum emission is dominated by the outer regions of the BH accretion 
disc, or by a young star cluster around the BH \citep{sor12,far12}.

In the absence of phase-resolved dynamical measurements of the BH motion, 
we can use the {\it Swift} X-ray lightcurve
properties to constrain the system parameters.
The X-ray flux shows recurrent outbursts every $\approx (366 \pm 4)$ d 
(seen every late August in 2009, 2010, 2011 and 2012), due either to some kind 
of disc instability, or to a periodic enhancement 
of the accretion rate. Several alternative scenarios were considered and 
discussed by \citet{las11}, who favoured a model in which enhanced mass 
transfer into a quasi-permanent accretion disc is triggered by the passage 
at periastron of an asymptotic giant branch (AGB) star on an eccentric 
orbit ($e \sim 0.7$).
Since the publication of that work, the detection of the third and fourth 
consecutive outbursts (see \citealt{god12} for the first report of 
this year's outburst) 
has clinched the interpretation of the recurrence 
timescale as the binary period. Furthermore, additional optical photometric 
results have been published, 
based on data from the {\it Hubble Space Telescope (HST)} \citep{far12}
and from the European Southern Observatory (ESO)'s Very Large Telescope (VLT)
\citep{sor12}. Thus, in this paper we revisit and update \citet{las11}'s 
orbital models and constraints in the light of the new results.


\section{Size of the accretion disc}

\subsection{Predictions for a standard disc model}

At a distance of 95 Mpc, the characteristic size of the region responsible 
for most of the soft, thermal ($kT \approx 0.2$ keV) X-ray emission  
is $\sim$ a few $10^9$ cm
(inferred from fits to {\it XMM-Newton}, {\it Chandra}, {\it Swift} spectra), 
and is consistent with being constant 
during the decline of individual outbursts, and over the three 
recorded outbursts \citep{far09,ser11,dav11,sor11,sor12,far12}.
This suggests that the soft X-ray emission traces the true inner radius 
of the disc, bounded by the innermost stable circular orbit around the BH.
Instead, much less is known about the outer disc radius, from 
UV/optical/IR observations; it is still debated  
how much of the blue optical emission comes from an irradiated disc, 
and how much from a possible cluster of young stars around the BH.
If the disc is the dominant UV/optical emitter, the {\it HST} and VLT studies 
of \citet{far12} and \citet{sor12}, respectively, agree on an outer disc 
radius $\approx 10^{13}$ cm $\sim 1$ au\footnote{Note for the arXiv version: 
after this paper was accepted in MNRAS, a re-analysis of the {\it HST} data 
with a different disc irradiation model, by Mapelli et al. 
(submitted to MNRAS), showed that they are consistent 
with an outer disk radius $\approx 3.5 \times 10^{13}$ cm, for a viewing 
angle $i = 45^{\circ}$; the radius is smaller for a face-on view, 
which we consider a more likely scenario, given the narrow 
full-width-half-maximum of the H$\alpha$ emission line (work by Hau \& Soria, 
in preparation). Even if we adopt this slightly larger 
upper limit for the disc radius, the argument for a high eccentricity 
discussed in Section 3 remains unchanged.
}. 
If a substantial contribution comes 
from unresolved young stars, we can take that value as an upper limit 
to the true disc size.
A ratio of outer/inner disc radii $\sim 10^3$ is significantly smaller 
than observed in transient Galactic BHs with Roche-lobe-filling donors, 
where typical outer radii are $\sim 10^{11}$ cm $\sim$ a few $10^4$ times 
the innermost stable circular orbit \citep{zur11,hyn02,hyn98}. 
This serves as a warning that 
we have to disentangle what scales with BH mass and what does not, when using 
scaled-up Galactic BH models to interpret HLX-1. While the inner disc 
depends directly on the BH mass, the outer disc depends mostly on the donor 
star and binary separation.

There is an alternative way to estimate the outer disc size, based 
on the X-ray outburst decline timescale. 
Following \citet{kin98} and \citet{fra02}, we assume that the outbursting disc 
is approximately in a steady state with surface density
\begin{equation}
\Sigma \equiv \rho H \approx \frac{\dot{M}_{\rm BH}}{3\pi \nu}
\end{equation}
where $\dot{M}_{\rm BH}$ is the central accretion rate and $\nu$ 
the kinematic viscosity.
When the whole disc from $R_{\rm in}$ to $R_{\rm out}$ is in a hot, 
high-viscosity state, the total mass in the disc
\begin{equation}
M_{\rm disc} = 2\pi \int_0^{R_{\rm out}} \Sigma R\,dR 
   \approx \frac{\dot{M}_{\rm BH} R^{2}_{\rm out}}{3\nu} 
   = \frac{-\dot{M}_{\rm disc} R^{2}_{\rm out}}{3\nu},
\end{equation}
where we have neglected other sources of mass loss from the disc apart 
from BH accretion. In Eq.(2), $\nu$ is interpreted as an average value 
of the kinematic viscosity over the whole disc; in practice, 
we take the value of $\nu$ near the outer edge of the disc \citep{kin98}.
Integrating Eq.(2), we obtain the well-known exponential decline 
for the disc mass
\begin{equation}
M_{\rm disc} = M_{\rm{disc},0} \exp(-3\nu t/R^{2}_{\rm out}),
\end{equation}
and consequently also for the accretion rate
\begin{equation}
\dot{M}_{\rm BH} = \frac{3\nu M_{\rm{disc},0}}{R^{2}_{\rm out}} 
     \exp(-3\nu t/R^{2}_{\rm out}),
\end{equation}
and the outburst luminosity $L \sim L_{\rm X} \sim 0.1\dot{M}_{\rm BH}c^2$. 
In summary, we expect to see a luminosity  
\begin{equation}
L_{\rm X} \approx L_{{\rm X},0} \exp(-3\nu t/R^{2}_{\rm out}),
\end{equation}  
where $L_{{\rm X},0}$ is the value at the outburst peak, declining 
on a timescale 
\begin{equation}
\tau_{\rm e} \approx R^{2}_{\rm out}/(3\nu),
\end{equation}
as long as the rate at which the disc mass is depleted during 
the outburst decline is much larger 
than any ongoing transfer of mass from the donor star.
For the viscosity at the outer edge of the disc, we take 
the usual parameterization $\nu = \alpha c_{\rm s} H$ \citep{sha73},
where $\alpha$ is the viscosity coefficient in the hot state, 
$c_{\rm s}$ is the sound speed and $H$ the vertical scaleheight. 
In the simplest, order-of-magnitude approximation, we can take 
an outer disc temperature $\approx 10^4$ K (enough to keep it 
in the hot state), corresponding to $c_{\rm s} \approx 2 \times 10^6$ 
cm s$^{-1}$, and a vertical height $H \approx 0.1 R$.
This gives, from Eq.(6):
\begin{equation}
R_{\rm out} \sim 6 \times 10^5 \, \alpha \, \tau_{\rm e} \ {\rm cm},
\end{equation}
which we shall directly compare with the observations.

If we adopt the Shakura-Sunyaev disc solution \citep{sha73,fra02}, 
with Kramers opacity, we can write 
\begin{eqnarray}
\nu & \approx & 1.8 \times 10^{14} 
  \alpha^{4/5} \dot{M}^{3/10}_{16} m_1^{-1/4} R^{3/4}_{10} 
  \ {\mathrm {cm}}^2 {\mathrm s}^{-1} \nonumber\\
& \approx & 6.4 \times 10^{16} \, 
\alpha^{4/5} \dot{M}^{3/10}_{22} m_3^{-1/4} R^{3/4}_{12} 
  \ {\mathrm {cm}}^2 {\mathrm s}^{-1},
\end{eqnarray}
where $\dot{M}_{16}$ is the accretion rate in units of $10^{16}$ g s$^{-1}$, 
$m_1$ is the BH mass in solar units, 
$R_{10} \equiv R_{\rm out}/(10^{10} \rm{cm})$, etc.
Then, from Eq.(6):
\begin{equation}
\tau_{\rm e} \approx 5.2 \times 10^6 \alpha^{-4/5} \dot{M}^{-3/10}_{22} m_3^{1/4} 
     R^{5/4}_{12} \ {\mathrm s},
\end{equation}
that is 
\begin{equation}
R_{12} \approx \left(\frac{\tau_{\rm e}}{5.2 \times 10^6}\right)^{4/5} \, 
    \alpha^{16/25} \dot{M}^{6/25}_{22} m_3^{-1/5}.
\end{equation}

The peak luminosity $L_{{\rm X},0}$ in Eq.(2) can be left as a purely 
observational parameter, or it can itself be expressed as a function  
of outer disc radius, viscosity, density and BH mass, if we assume that 
the outburst is triggered via the dwarf-nova instability 
\citep{can93}. (More precisely, if we assume that the outburst starts when 
the enhanced mass transfer due to periastron passage pushes the disk 
from the cold to the hot state.) 
In that case, the surface density at any radius 
immediately before the start of the outburst approaches 
the maximum value allowed by the S-curve in the surface-temperature 
phase space \citep{kin98,fra02}:
\begin{eqnarray}
\Sigma_{\rm max} & \approx & 11.4 R_{10}^{1.05} m_1^{-0.35} \alpha_{\rm c}^{-0.86} 
    \ {\rm{g}} \  {\rm{cm}}^{-2} \nonumber\\
    & \approx & 1.3 \times 10^2 R_{12}^{1.05} m_3^{-0.35} \alpha_{\rm c}^{-0.86} 
    \ {\rm{g}} \  {\rm{cm}}^{-2}, 
\end{eqnarray}
where $\alpha_{\rm c} \sim 0.01$ is the viscosity parameter in the cold 
disc state. Taking for simplicity $H \approx bR$, where $b$ is a constant 
$\sim 0.1$, we can then express the maximum volume density at the start 
of the outburst as $\rho_{\rm max} = \Sigma_{\rm max}/H \approx 
\Sigma_{\rm max}/(bR)$, which is essentially independent of $R$, given 
the expression for $\Sigma_{\rm max}$ in Eq.(11).
It is then easy to integrate the total disc mass at the start 
of the outburst:
\begin{eqnarray}
M_{\rm{disc},0} &=& 2\pi \int_0^{R_{\rm out}} \Sigma_{\rm max} R\,dR \nonumber \\
     &\approx& (2\pi b) \int_0^{R_{\rm out}} \rho R^2\,dR
      = (2\pi b) \frac{\rho R_{\rm out}^3}{3}
\end{eqnarray}
the disc mass at later times
 \begin{equation}
M_{\rm{disc}} \approx 
          \frac{(2\pi b) \rho R_{\rm out}^3}{3} \, \exp(-3\nu t/R^{2}_{\rm out}),
\end{equation}
the accretion rate (cf. Eq.(4))
\begin{equation}
\dot{M}_{\rm BH} \approx (2\pi b) (R_{\rm out} \nu \rho) \, 
        \exp(-3\nu t/R^{2}_{\rm out}),
\end{equation}
and the peak luminosity 
\begin{eqnarray}
L_{{\rm X},0} & \approx& (0.1 c^2) (2\pi b) (R_{\rm out} \nu \rho)  \nonumber\\
    &\approx & (0.1 c^2) (2\pi b) R_{\rm out} (\alpha c_{\rm s} H) 
                 (\Sigma_{\rm max}/H) 
          \nonumber\\
    &\approx & 0.1 \alpha c^2 (2\pi b) R_{\rm out} c_{\rm s} \Sigma_{\rm max},
\end{eqnarray}
where $\Sigma_{\rm max}$ comes from Eq.(11).

So far, we have assumed that the whole disc is in the hot state; 
this is usually the case in the early part of an outburst, especially  
when the outer edge of the disc is kept in the hot state by X-ray irradiation. 
The exponential decay continues until the outer disc annuli can no longer 
be kept in the hot state, so that hydrogen recombines and viscosity drops. 
From that moment, the outer edge 
of the hot disc $R_{\rm h} < R_{\rm out}$. The contribution 
to the accretion rate and to the continuum 
X-ray/UV/optical emission from the outer (cold, low-viscosity) annuli 
at $R_{\rm h} < R < R_{\rm out}$ becomes negligible.
It was shown by \citet{kin98} that the central accretion rate 
in this second phase of the decline is
\begin{equation}
\dot{M}_{\rm BH} = \dot{M}_{\rm BH} (t_1) \left[1 - C (t-t_1)\right],
\end{equation}
where $t_1$ is the time after which the outer disc is no longer 
in the hot state, and 
$C$ parameterizes the fraction of X-ray luminosity 
intercepted and thermalized in the outer disc. As $C$ can be taken 
as a constant, Eq.(16) shows that the accretion rate 
and luminosity decline in the late part of the outburst is linear.
Most importantly for our current purpose, the slope 
of the linear decline is such that 
\begin{equation}
t_{\rm end} - t_1 = \tau_{\rm e}
\end{equation}
\citep{kin98}, where $t_{\rm end}$ is the (extrapolated) time in which 
the accretion rate and luminosity go to zero.

Finally, we need to consider the case when there is ongoing mass 
transfer $\dot{M}_2$ from the donor star during the outburst. In that case, 
the asymptotic value of the luminosity in the exponential decline 
is not zero but $L_2 \approx 0.1 (-\dot{M}_2) c^2$ (assuming a standard 
radiative efficiency $\approx 0.1$). Recalling that $t_1$ 
is the time when the lightcurve switches from an exponential to a linear 
decline, and defining $L_1 \equiv L(t_1)$, Eq.(5) is modified as \citep{pow07}
\begin{equation}
L_{\rm X} = \left(L_1 - L_2\right) 
   \exp(-3\nu \left(t-t_1\right)/R^{2}_{\rm out}) + L_2.
\end{equation}
After the transition to a linear regime, the luminosity is 
\begin{equation}
L_{\rm X} = L_1 \left[1 - \frac{3\nu}{R^{2}_{\rm out}} \left(t - t_1\right)\right].
\end{equation}
Note that if $-\dot{M}_2 > 0$, the first derivative of the luminosity 
is discontinuous at $t=t_1$ \citep{pow07}, because the gradient 
of the exponential decay is 
\begin{equation}
\dot{L}_{\rm X}(t_1) \approx  - \frac{3\nu}{R^{2}_{\rm out}} \, L_1 \,
    \left(1+ \frac{0.1 \dot{M}_2 c^2}{L_1}\right),
\end{equation}
that is flatter than the gradient of the linear decay 
\begin{equation}
\dot{L}_{\rm X}(t_1) = - \frac{3\nu}{R^{2}_{\rm out}} \, L_1.
\end{equation}
Therefore, the exponential-to-linear transition is often referred 
to as the ``knee'' in the lightcurve of transient X-ray binaries.

\subsection{Comparison with the observations}

We shall now fit the X-ray lightcurve to obtain two independent estimates 
of the viscous timescale $\tau_{\rm e}$, from the exponential and 
the linear regime, and use them to constrain $R_{\rm out}$  
from Eq.(10), or, using a simpler approximation for the scale-height, 
from Eq.(7). We shall then derive an independent estimate of $R_{\rm out}$ 
from the expression for the peak luminosity in Eq.(15).
We studied the publicly 
available\footnote{http://www.swift.ac.uk/user\_objects} 
{\it Swift} X-Ray Telescope \citep[XRT;][]{bur05} data  
for the 2010 outburst, because it is the same outburst for which we obtained 
constraints on $R_{\rm out}$ from the optical continuum \citep{sor12,far12}.
We used the on-line {\it Swift}/XRT data product generator\footnote{
	Including the new treatment of the vignetting correction,
	introduced after 2011 August 5.}
   \citep{eva07,eva09}
   to extract a lightcurve in the $0.3$--$10$~keV band.
We fitted the lightcurve with an initial exponential decay, a knee and 
a linear decay (Fig.~1).
The shape of the X-ray outburst lightcurve of HLX-1 is remarkably 
similar to those of several 
transient Galactic X-ray binaries (BHs and neutron stars), modelled 
by \citet{pow07}, which were successfully used to constrain the size 
of their accretion discs.

\begin{figure}
\psfig{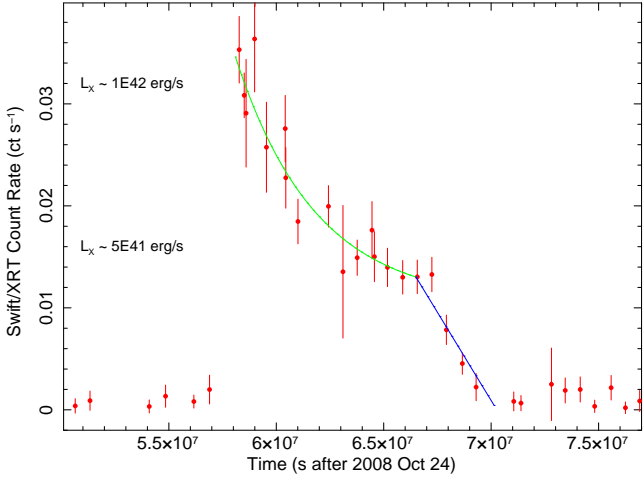}
\caption{{\it Swift}/XRT lightcurve of the 2010 outburst, fitted with 
a standard X-ray transient model (exponential decay, knee, linear decay). 
}
\label{f1}
\end{figure}

For the exponential part, we obtain a best-fitting timescale 
$\tau_{\rm e} = R^{2}_{\rm out}/(3\nu) = 3.7^{+5.0}_{-1.5} \times 10^6$ s 
($90\%$ confidence limit). 
For the linear part, we have $\tau_{\rm e} = 3.5^{+1.0}_{-0.8} \times 10^6$ s.
Assuming a peak luminosity $\approx 10^{42}$ erg s$^{-1}$ 
in the $0.3$--$10$ keV band, 
and a bolometric luminosity a factor of 2 higher, implies 
an accretion rate $\dot{M} \approx 2\times 10^{22}$ g s$^{-1}$ 
at standard efficiency. The viscosity parameter $\alpha \la 1$, and more likely 
$\alpha \sim 0.3$ \citep{fra02}. From Eq.(10), this implies 
$R_{\rm out} \approx 10^{12}$ cm, only very weakly dependent on BH mass 
and accretion rate. Using the approximation in Eq.(7), we also 
obtain $R_{\rm out} \sim 10^{12}$ cm. 
From the peak luminosity (Eq.(15)), for $\alpha \sim 0.3$ 
we obtain $R_{\rm out} \sim 4 \times 10^{12}$ cm.

We also analyzed the lightcurves for the 2009 and 2011 outbursts 
(the 2012 outburst was still ongoing as this paper went to press).
They are more noisy, less easy to interpret in terms 
of exponential and linear branches. However, for both of them 
it is possible to estimate an e-folding decline timescale, 
roughly corresponding 
to the exponential timescale determined for the 2010 outburst. 
The timescales are $\approx 5 \times 10^6$ s and $\approx 3 \times 10^6$ s 
for 2009 and 2011, respectively, and the peak 
luminosities are approximately the same in all three outbursts. 
Thus, we also estimate an outer radius $\sim 10^{12}$ cm 
in the 2009 and 2011 outbursts, in the standard disc approximation.

Those values are almost one order of magnitude smaller than what we estimated 
by assuming that most of the continuum emission comes from the hot disc; 
and the latter was already a surprisingly small radius compared with 
the binary system parameters (see Section 3).
We note that both the {\it HST} and VLT observations 
\citep{sor12,far12} were 
taken during the exponential part of the decline, that is when the whole disc 
was in a hot state. Therefore, the estimates of $R_{\rm out}$ 
from the optical/UV continuum should be comparable to those from 
the X-ray lightcurve. New optical observations early in the next 
outburst will hopefully allow us to measure the true outer disc 
size and luminosity.


\section{Orbital parameters}

We can now compare the size of the disc estimated from optical and X-ray 
flux measurements ($10^{12}$ cm $\la R_{\rm out} \la 10^{13}$ cm) with 
the characteristic size of the binary system. 
Because of the sharpness of the outbursts rise and decline, 
reminiscent of Galactic X-ray binaries, we assume that 
the BH is accreting from a single donor star rather than AGN-like gas inflows. 
We also assume that the outburst recurrence timescale $\sim 370$ d 
corresponds to the binary period.
Then, the semimajor axis $a$ of the binary is 
\begin{equation}
a = 1.50 \times 10^{13} m^{1/3} (1+q)^{1/3} P_{\rm yr}^{2/3} 
    \ {\rm {cm}}
\end{equation}
\citep{new87}, where $q = M_2/M_{\rm BH}$ 
and $m \equiv (M_{\rm BH} + M_2)/M_{\odot} \equiv M/M_{\odot}$. 
Typical values for HLX-1 in the intermediate-mass BH scenario are 
$q \sim 10^{-3}$ and $m^{1/3} \sim 10$--$20$. Therefore, 
in the most accepted scenario, the semimajor axis is at least 10, 
and possibly up to 100 times larger than the disc radius. 
This mismatch clearly suggests an eccentric orbit \citep{las11}, 
in which the characteristic disc size is determined by the periastron 
separation $R_{\rm per} = (1-e) a$, with eccentricity $e \ga 0.9$.

The amount of mass transferred to the BH in each outburst suggests that 
the donor star overflows its instantaneous Roche lobe every time it passes 
at periastron. In Roche-lobe mass transfer systems, the outer edge 
of the accretion disc is generally identified with the largest stable 
non-intersecting orbit (tidal truncation radius $R_{\rm T}$). For mass ratios 
$M_2/M_1 \ll 1$, $R_{\rm T} \approx 0.48 a$ \citep{pac77,pap77,whi88,war95}. 
If unstable orbits are also allowed, the disc may expand up to 
$R_{\rm T} \approx 0.60a/(1+q)$ 
\citep[][and references therein]{war95}.
If we take the periastron distance as the instantaneous binary separation 
during the phase of Roche-lobe overflow, this corresponds to 
an expected disc size $R_{\rm out} \approx 0.6 (1-e) a$. 
For an observed disc size $R_{\rm out} = 10^{13}$ cm, the tidal radius condition 
would require an eccentricity $e \approx 0.89$ for a BH mass $= 1000 M_{\odot}$, 
and $e \approx 0.95$ for a BH mass $= 10^4 M_{\odot}$.
If we take the lower bound to our observed disc size $R_{\rm out} = 10^{12}$ cm, 
we need $e \approx 0.989$ or $e \approx 0.995$, respectively.


We can argue that the tidal truncation constraint is not relevant 
to the case of HLX-1, where mass transfer may occur impulsively near 
periastron, and the timescale for the disc to expand to its tidal truncation 
radius is similar to the timescale for the disc matter to be accreted 
and for the binary orbit to expand after periastron. 
In other words, in HLX-1 the disc may look small because it did not have 
time to grow to its tidal truncation radius. 
Instead, the circularization radius provides a stronger 
lower limit to the predicted disc size, and is applicable to any system 
where mass transfer occurs through the Lagrangian point L$_1$.

The circularization radius $R_{\rm cir}$ is defined via the conservation 
of angular momentum equation 
\begin{equation}
v_{\phi}\left(R_{\rm cir}\right) R_{\rm cir} = \left(X_{\rm L1} R_{\rm per}\right)^2 
\Omega \left(R_{\rm per}\right),
\end{equation}
where $v_{\phi}$ is the orbital velocity of the accretion stream around the BH, 
$X_{\rm L1}R_{\rm per}$ the distance between the BH and the Lagrangian point 
L$_1$, and 
$\Omega \left(R_{\rm per}\right)$ the angular velocity of the donor star 
at periastron. Here, we must be careful not to use the well-known fitting 
formula for $X_{\rm L1} \approx 0.500 -0.227 \log q$ \citep{fra02} because 
it applies only for $q > 0.1$ and for circular orbits. Instead, we need 
to compute $X_{\rm L1}$ from Eq.(A13) of \citet{sep07}, valid 
for eccentric orbits:
\begin{equation}
\frac{q}{\left(1-X_{\rm L1}\right)^2} - \frac{1}{X_{\rm L1}^2} 
   -f^2 \left(1-X_{\rm L1}\right) (1+q)(1+e) + 1 = 0.
\end{equation}
Here, $f$ is the ratio of the rotational angular velocity 
of the donor star to its orbital angular velocity, at periastron. 
It parameterizes the degree of tidal locking; a star 
with $f=1$ is rotating synchronously with the orbit, at periastron. 
A list of $X_{\rm L1}$ solutions 
for characteristic values of $q$ and $e$ is given in Table 1.
The orbital velocity of the accretion stream around the BH 
\begin{equation}
v_{\phi} \approx \left(\frac{GM_{\rm BH}}{R_{\rm cir}}\right)^{1/2}
\end{equation}
and the orbital angular velocity at periastron 
\begin{equation}
\Omega\left(R_{\rm per}\right) = \frac{(1+e)^{1/2}}{(1-e)^{3/2}} \, 
  \left[\frac{G\left(M_{\rm BH}+M_2\right)}{a^3}\right]^{1/2}.
\end{equation}
For the sake of our numerical estimate, we take $X_{\rm L1} = 0.95$, 
a typical value in the range of parameters thought to be relevant for HLX-1 
(Table 1).
Then, substituting into Eq.(23), we have:
\begin{eqnarray}
R_{\rm cir} &\approx& \frac{\left(0.95 R_{\rm per}\right)^4}{GM_{\rm BH}} \, 
    \frac{(1+e)}{(1-e)^3} \, \frac{G\left(M_{\rm BH} + M_2\right)}{a^3} 
   \nonumber \\
      &\approx& 0.81 (1-e^2) (1+q) a   \nonumber \\
         &\approx& 1.2 \times 10^{14} m_3^{1/3} (1-e^2) (1+q)^{4/3} 
          P^{2/3}_{\rm yr} \  {\rm {cm}}.
\end{eqnarray}
Assuming that the BH mass is in the range $\sim 10^3$--$10^4 M_{\odot}$, 
Eq.(27) gives us a strong constraint on $e$, by imposing that 
the circularization radius is smaller than the observed outer disc size.
For example, assuming $M_{\rm BH} = 5 \times 10^3 M_{\odot}$, a circularization 
radius $R_{\rm cir} = 10^{13}$ cm (at the upper end of our disc size estimates) 
requires $e \approx 0.97$; for 
$R_{\rm cir} = 10^{12}$ cm (at the lower end of our disc size estimates), 
$e \approx 0.997$.
The corresponding periastron distances\footnote{Note that for $e\ga 0.2$, 
the periastrion distance between the two stars is always smaller 
than the circularization radius around the accreting primary. This is  
because the angular momentum of the secondary at periastron, 
and of the matter transiting through the L$_1$ point,  
is larger than the angular momentum of a circular orbit with the same 
binary separation. It is not a problem, because by the time the 
accretion stream has completed a full orbit around the BH and formed 
a ring, the secondary has moved away from periastron and the primary's 
Roche lobe has widened.} 
$R_{\rm per} = (1-e) a \approx 6 \times 10^{12}$ cm for $e=0.97$, and 
$R_{\rm per} \approx 6 \times 10^{11}$ cm for $e=0.997$.

\begin{table}
\begin{center}
\begin{tabular}{lcccc}\hline
\multicolumn{5}{c}{f=0} \\
\hline
 &  $e=0.80$ & $e=0.90$ & $e=0.95$ & $e=0.99$\\
\hline\\[-5pt]
$q=10^{-1}$ & \multicolumn{4}{c}{$0.695$} \\
$q=10^{-2}$ & \multicolumn{4}{c}{$0.843$} \\
$q=10^{-3}$ & \multicolumn{4}{c}{$0.924$} \\
$q=10^{-4}$ & \multicolumn{4}{c}{$0.964$} \\
\hline
\multicolumn{5}{c}{f=1} \\
\hline\\[-5pt]
$q=10^{-1}$ & $0.732$ &    $0.734$ &    $0.735$ &    $0.735$ \\
$q=10^{-2}$ & $0.868$ &    $0.869$ &    $0.869$ &    $0.869$ \\
$q=10^{-3}$ & $0.937$ &    $0.938$ &    $0.938$ &    $0.938$ \\
$q=10^{-3}$ & $0.970$ &    $0.971$ &    $0.971$ &    $0.971$ \\
\hline
\end{tabular} 
\end{center}
\caption{Distance $X_{\rm L1}$ between the BH and the L$_1$ point 
(as a fraction of the periastron distance $R_{\rm per}$), 
in the parameter range of interest for HLX-1; from Eq.(24).}
\label{tab1}
\end{table}

The distance between L$_1$ and the centre of the donor star 
places an upper limit on its radius; characteristic values 
of its instantaneous volume-averaged Roche lobe radius at periastron  
can be obtained from \citet{egg83}. For typical 
values $q \sim 10^{-4}$--$10^{-3}$, the secondary gets squeezed 
to a radius $\approx (0.02$--$0.05)R_{\rm per}$. 
Since we have assumed that the secondary fills the Roche lobe and 
dumps mass into the BH only near periastron, 
this radius must also be similar to the size of the donor star.
For example, if $q = 2\times 10^{-4}$ and $e =0.97$, the donor star 
must have a radius $\approx 4 R_{\odot}$, consistent with main sequence 
and subgiant stars.

The values of $e$ estimated here from disc size arguments 
are much more extreme than what was suggested in \citet{las11}.
They may seem implausible, knowing that tidal forces tend 
to circularize orbits in X-ray binaries.
However, \citet{sep07,sep09} showed that in the case 
of a donor star that transfers mass impulsively only at periastron, 
with $q \la 1 -0.4e +0.18e^2$, the secular evolution of the orbit leads 
to an increase of both eccentricity 
and semimajor axis, even when the opposite effect 
of tidal forces is taken into account.
If tidal circularization is neglected, 
the eccentricity increases as
\begin{equation}
\langle\dot{e}\rangle = \frac{1}{\pi} \frac{\langle\dot{M_2}\rangle}{M_2} 
    (1-e^2)^{1/2} (1-e)(q-1)
\end{equation}
and the semimajor axis (and hence the binary period) as
\begin{equation}
\langle\dot{a}\rangle = \frac{a}{\pi} \frac{\langle\dot{M_2}\rangle}{M_2} 
    (1-e^2)^{1/2}(q-1).
\end{equation}
It is then easy to show from Eq.(28,29) that the periastron distance 
$R_{\rm per} = (1-e)a$, and therefore also the size of the secondary's 
Roche lobe at periastron, remains unchanged. 

Finally, we need to assess what donor stars can survive on such 
eccentric orbits with small periastron distance, avoiding 
tidal disruption. The condition for survival is that 
the periastron distance $R_{\rm per}$ is larger than 
the tidal disruption radius \citep{ree88}
\begin{equation}
R_{\rm td} \approx 5 \times 10^{11} m_3^{1/3} \left(R_2/R_{\odot}\right) 
   \left(M_2/M_{\odot}\right)^{-1/3} \ {\rm {cm}}.
\end{equation}
By substituting $R_2 \la 0.05 R_{\rm per}$ into Eq.(31), we can re-cast the 
tidal survival condition as
\begin{equation}
M_2 \ga 4.6 \times 10^{-5} M_{\rm BH}, 
\end{equation}
easily satisfied in the likely mass range of HLX-1.


\section{Conclusions}

We compared the estimates of the disc size from the optical continuum flux 
($R_{\rm out} \la 10^{13}$ cm) with those obtained by fitting 
the X-ray luminosity decline after an outburst. 
We found that, at least for the 2010 outburst, 
the decline displays the standard sequence of exponential phase, 
knee, linear phase often seen in Galactic X-ray binaries; this strengthens 
the interpretation that the thermal X-ray emission in HLX-1 comes from a disc, 
and the decline timescale corresponds to its viscous timescale.
For all three outbursts observed to-date, the timescale 
is very short, $\sim 4$--$8$ weeks. This is similar or only slightly longer 
than what is typically observed in Galactic X-ray transients, 
despite that fact that both the orbital period and the BH mass 
(and, hence, the semimajor axis) of HLX-1 
are $\ga 100$ times larger. The outer disc radius estimated from 
the viscous timescale is $R_{\rm out} \sim 10^{12}$ cm, 
if the viscosity parameter is similar to the values usually 
estimated for Galactic BH transients in a high state.
We cannot rule out that the fast accretion of the disc matter in HLX-1 
may be partly due to an effective viscosity $\alpha_{\rm eff} \ga 1$, 
higher than in the Shakura-Sunyaev prescription. But we argue 
that even if we assume the upper disc size estimate 
$R_{\rm out} \sim 10^{13}$ cm, 
a highly eccentric orbit is required to explain the small disc size.

To quantify the eccentricity, we calculated the characteristic length-scales 
of the binary system, as a function of BH mass and eccentricity.
If the disc extends at least as far as the circularization radius 
(as is usually the case in X-ray binaries with Roche-lobe mass transfer), 
we obtain that $R_{\rm cir} \sim (1-e^2)a$, and therefore
$e \ga 0.95$ for a BH mass $\ga 10^3 M_{\odot}$.
We argued that X-ray binaries with such extreme values of $e$ 
are the most likely evolutionary endpoint of systems with $q \ll 1$ 
and a moderately eccentric initial orbit, such that Roche-lobe-overflow 
mass transfer occurs only impulsively near periastron. Secular evolution 
will tend to make the orbit more and more eccentric, by increasing 
the semimajor axis and the binary period, at constant periastron distance.

The small periastron distance required to explain 
the HLX-1 observations sets an upper limit to the current radius 
of the donor star $R \la$ a few $R_{\odot}$, ruling out supergiants, 
red giants and AGB stars. Possible donors are main sequence 
(B type or later) or subgiants.
The compactness of the donor star, and the fact that secular orbital 
evolution due to mass transfer will not change the periastron distance, 
imply that the companion star in HLX-1 is not at immediate risk 
of tidal disruption, and will not be in the near future. 
In other words, we are not observing 
HLX-1 in a peculiar moment of its evolution, immediately prior 
to tidal break-up of the donor star. HLX-1 appears to be a stable 
system, with a lifetime for X-ray outbursts determined primarily by 
the mass transfer timescale from the donor; at a rate $\sim 10^{-5} M_{\odot}$  
yr$^{-1}$ (averaged over the binary period), it may last for another 
$\sim 10^5$--$10^6$ yr, during which its semimajor axis and binary 
period (and hence, interval between outbursts) will continue 
to increase.

Eccentricities $\ga 0.95$ may seem implausibly extreme, 
but there is at least one class of stellar objects where they are the norm:
S stars observed on highly eccentric orbits within 0.01 pc of the Galactic 
nuclear BH \citep{ale05,gil09}. A possible scenario for the origin 
of Galactic S stars is the tidal disruption of a stellar binary system 
near the BH, which produces an escaping, hyper-velocity star, and 
a more tightly bound star on a very eccentric orbit, theoretically as high 
as $e \approx 0.99$ \citep{loc08}. Observationally, the most eccentric, 
bound S star for which orbital parameters have been reliably determined 
has $e \approx 0.96$ \citep{gil09}. We speculate that intermediate-mass 
BHs in star clusters may also capture stellar companions 
on very eccentric orbits through a similar process.

\section*{Acknowledgments}

I thank Guillaume Dubus, Sean Farrell, Jeanette Gladstone, 
Pasi Hakala, Michela Mapelli, George Hau, Albert Kong, Tom Russell, 
Luca Zampieri for insightful discussions 
on the nature of HLX-1. This work was improved also thanks to the feedback 
received from several colleagues after my presentation 
at the IAU Symposium 290 in Beijing.


\end{document}